\documentclass[conference, letterpaper]{IEEEtran}
\IEEEoverridecommandlockouts

\usepackage{cite}
\usepackage{bm}
\usepackage{amsmath,amssymb,amsfonts}
\usepackage{algorithm}
\usepackage{algorithmic}
\usepackage{graphicx}

\usepackage{graphicx}
\usepackage{subcaption}

\usepackage[letterpaper, left=0.625in, right=0.625in, bottom=1.02in, top=0.70in]{geometry}

\usepackage{textcomp}
\usepackage{xcolor}

\newcommand{\be}{\begin{equation}}
\newcommand{\ee}{\end{equation}}

\def\BibTeX{{\rm B\kern-.05em{\sc i\kern-.025em b}\kern-.08em
    T\kern-.1667em\lower.7ex\hbox{E}\kern-.125emX}}

\begin{document}

\title{A New Implementation of Federated Learning for Privacy and Security Enhancement
}
\author{\IEEEauthorblockN{Xiang Ma\IEEEauthorrefmark{1}, Haijian Sun\IEEEauthorrefmark{2}, Rose Qingyang Hu\IEEEauthorrefmark{1} and Yi Qian\IEEEauthorrefmark{3}}
\IEEEauthorblockA{
\IEEEauthorrefmark{1}Department of Electrical and Computer Engineering, Utah State University, Logan, UT \\
\IEEEauthorrefmark{2}Department of Computer Science, University of Wisconsin-Whitewater, Whitewater, WI \\
\IEEEauthorrefmark{3}Department of Electrical and Computer Engineering, University of Nebraska–Lincoln, Omaha, NE \\
Emails: \IEEEauthorrefmark{1}\{xiang.ma@ieee.org, rose.hu@usu.edu\}, \IEEEauthorrefmark{2}h.j.sun@ieee.org
, \IEEEauthorrefmark{3}yi.qian@unl.edu}
}
\maketitle

\begin{abstract}
Motivated by the ever-increasing concerns on personal data privacy and the rapidly growing data volume at local clients, federated learning (FL) has emerged as a new machine learning setting. An FL system is comprised of a central parameter server and multiple local clients. It keeps data at local clients and learns a centralized model by sharing the model parameters learned locally. No local data needs to be shared, and privacy can be well protected. Nevertheless, since it is the model instead of the raw data that is shared, the system can be exposed to the poisoning model attacks launched by malicious clients. Furthermore, it is challenging to identify malicious clients since no local client data is available on the server. Besides, membership inference attacks can still be performed by using the uploaded model to estimate the client's local data, leading to privacy disclosure. In this work, we first propose a model update based federated averaging algorithm to defend against Byzantine attacks such as additive noise attacks and sign-flipping attacks. The individual client model initialization method is presented to provide further privacy protections from the membership inference attacks by hiding the individual local machine learning model. When combining these two schemes, privacy and security can be both effectively enhanced. The proposed schemes are proved to converge experimentally under non-IID data distribution when there are no attacks. Under Byzantine attacks, the proposed schemes perform much better than the classical model based FedAvg algorithm.
\end{abstract}


\section{Introduction}
Federated learning (FL) \cite{federated} aims to build a robust machine learning (ML) model where local clients (LCs) distributively train their ML model using their locally collected data. In a typical FL setting, a central parameter server (PS) is connected to multiple clients and aggregates the models uploaded by the LCs. User privacy can be greatly protected since no local data is shared. However, it could be more harmful to the system when the malicious LCs launch Byzantine attacks by sending the poisoning model to the PS, which can directly degrade the overall learning performance. Furthermore, privacy is protected but not guaranteed in FL since attackers can still infer the private data and some key parameters such as gradients from the model in a membership inference attack \cite{membership}.

The Byzantine attack launched by the malicious LCs aims to degrade the learning performance. The model poisoning attacks can directly reduce the local task execution accuracy at LCs. Furthermore, the individual model of a single LC is invisible over the air or at the server due to the secure aggregation or other encoding methods. Only the aggregated model is available. Therefore, it is very difficult to identify malicious LCs in such a scenario. There have been extensive works to defend against Byzantine attacks in the distributed ML and FL. Most works aim to accommodate the attacks and mitigate the adverse effects, such as Krum aggregation \cite{krum} and geometric median aggregation \cite{geometric}. Krum aggregation is based on majority and squared distance to select some representative clients as the benign and trusted clients and estimate the true center using their updates. This may result in a biased model, especially when the data among LCs is not independent and identically distributed (non-IID). Unlike the classical arithmetic averaging, the geometric median considers the compounding that occurs from time to time and mitigates the impacts of poisoning attacks. But this scheme needs to obtain an individual model, which may cause privacy disclosure in membership inference attacks. In \cite{learning_model}, rather than accommodating the attacks as mentioned above, the work identified the attackers and removed the model updates coming from the attackers. However,  it could be challenging to select the dynamic threshold used to determine the attackers.

Another method to provide enhanced privacy protection in FL is differential privacy (DP) \cite{differential}. A typical approach of DP is to add random noise to the model and hide the real model from the eavesdroppers. It does not affect the system performance since the Gaussian noise can be averaged out due to aggregation. Another way for privacy protection is to hide the individual model from the eavesdroppers, i.e.,  only the aggregated model is accessible. This can be achieved with the Secure Aggregation protocol (SecAgg) \cite{secAgg} or over-the-air computation (AirComp) \cite{over_the_air} in wireless communications. Since the PS only needs to know the aggregated model, hiding the individual model does not impact the system performance. 

In this work, we propose to use model update based (MUB) aggregation to defend against Byzantine attacks and enhance security. The individual client model initialization (ICMI) scheme is further used to enhance privacy protection. By combining the two techniques as MUB-ICMI, privacy and security are enhanced. To the best of our knowledge, this is the first paper to provide both security and privacy protection without changing the fundamental structure of the classical federated averaging algorithm FedAvg \cite{fedavg}. The critical contributions of the paper are summarized as follows. 
\begin{itemize}

\item Since model update distribution has a much smaller deviation than the model distribution in federated learning, model based (MB) aggregation in the classical FedAvg is replaced by the MUB aggregation in FL. MUB-FL is robust to Byzantine attacks such as additive noise attacks and sign-flipping attacks while still achieving good learning performance.

\item By initializing the individual models at LCs rather than initializing the model at the server and uploading the local models using SecAgg or AirComp, the ICMI scheme can effectively hide the LC models. This protects LCs from membership inference attacks. 

\item The MUB scheme and the ICMI scheme can be combined as MUB-ICMI to enhance both security and privacy. Local learning models are well protected during the entire learning process by performing model initialization at each LC and uploading LC model updates instead of uploading LC models directly.

\item The simulation results show that MUB, ICMI, and MUB-ICMI can achieve a similar level of performance to the classical MB FedAvg without any attacks. While under Byzantine attack scenarios, both MUB and MUB-ICMI schemes are robust against attacks while still achieving good performance.

\end{itemize}

The rest of the paper is organized as follows. Section II introduces the classical MB FedAvg algorithm. It also gives the motivations to apply MUB, ICMI, and MUB-ICMI schemes in FL. Section III presents the detailed algorithms for the three proposed schemes. Simulation results are given in Section IV. Finally, section V concludes the paper. 

\section{System Model}
\subsection{FL System Model}
FL aims to learn a central ML model without data sharing. This is achieved by sharing the local ML models trained at each LC. The system is normally comprised of a central parameter server (PS) connected by $K$ LCs. Each LC has its dataset, and the data can be non-IID across different LCs. Considering bandwidth restrictions, especially in a wireless setting, only a fraction $C$ of total clients are selected to participate in the FL process in each round. In the classical MB FedAvg algorithm, the selected clients perform local learning based on their local dataset. Each LC then uploads the updated local model to the server for aggregation. The server employs the updated local models to get the latest global model via arithmetic averaging and then sends it back to all the LCs. The process repeats until the model converges.

The objective function of the FL system can be defined as
\be \label{objective_1}
\min_{\bm{w} \in R^d} f(\bm{w}),
\ee
where $f(\bm{w})=\frac{1}{|D|}\sum_{i=1}^n f_i(\bm{w})$, $|D|$ is the size of the dataset $D$. $f_i(\bm{w})=\ell(\bm{x_i}, y_i ; \bm{w})$ is the loss function used to capture the error between the data sample $(\bm{x_i}, y_i)$ and the mapping made by model parameters $\bm{w}$. Since data is distributed among $K$ clients, the objective \eqref{objective_1} can be rewritten as

\be
f(\bm{w}) = \sum_{k=1}^K \frac{|D_k|}{|D|} F_k(\bm{w}),
\ee
where, $F_k(\bm{w})=\frac{1}{|D_k|}\sum_{i \in D_k}f_i(\bm{w})$, $D_k$ is the dataset on client $k$. In each FL round, the client updates the local model based on the local data and the recently received global model, that is, 
\be \label{eq:local_model}
w_t^k = w_t - \eta \nabla F_k(w_t).
\ee
Here $\nabla F_k(w_t)$ is the gradient of $ F_k(w_t)$, $\eta$ is the stochastic gradient descent (SGD) step size or learning rate, $w_t$ is the received global model at round $t$, $w_t^k$ is the local model at round $t$ on client $k$.

\begin{figure}[!ht]
	\includegraphics[width=3.5in]{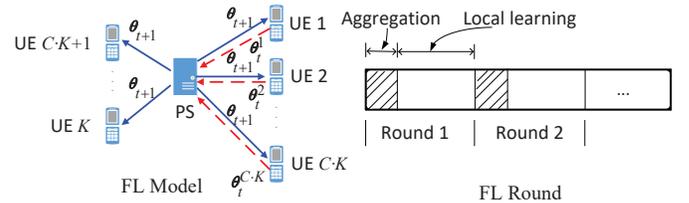}
	\centering
	\caption{FL Model \& FL Round}
	\label{model}
	\centering
\end{figure}

At round $t+1$, the server updates the global model as
\be \label{eq:aggregation}
w_{t+1} = \sum_{k=1}^K \frac{|D_k|}{|D|} w_t^k.
\ee
The FL learning repeats till the global model converges. A brief illustration of the FL model update is shown in Fig. \ref{model} left side. $\theta$ in Fig. \ref{model} represents the model parameter $w$ in the classical MB FedAvg algorithm. The FL learning process is shown on the right side. Each learning round consists of two parts. The first part (shaded area) is for model aggregation at the server, followed by local learning at each LC. Note the global model $w_t$ at  $t$ is the aggregation of the locally trained models at  $t-1$. The server initializes the global model in the first round. 

\subsection{Distribution of Model Update}

The simple arithmetic averaging algorithm (\ref{eq:aggregation}) in the classical MB FedAvg is easy to implement. However, the classical MB FedAvg is not robust under Byzantine attacks. Motivated by the difference between model distribution and model update distribution, we propose the MUB FedAvg algorithm in which each LC uploads the model update instead of the model itself to the server. The local model update is defined as the local model difference, i.e., 
\be
u_t^k=w_t^k - w_t^{k'},
\ee
where $w_t^k$ is defined in (3), i.e., the local model after the local learning process; $w_t^{k'}$ is the local model before local learning, which will be defined later in (7). The global model update is defined as 

\be
u_{t+1} = \sum_{k=1}^K \frac{|D_k|}{|D|}u_t^k.
\ee
The local ML model before local learning $w_t^{k'}$ is calculated based on the local model $w_{t-1}^k$  and the most recently received global model update $u_t$,
\be
w_t^{k'}=w_{t-1}^k+u_t.
\ee

At $t=1$, the server initializes the global model  $w_1$ as in the classical MB FedAvg and selects an initial global model update $u_1=0$. These two initial parameters are  broadcasted to LCs.  For convenience, $w_1$ in MUB-FL is referred to $w_0^k$. And explicitly, $w_0^k$ is the same for all LCs.  Please  note that $\theta$ in Fig. \ref{model} represents the model update $u$ in MUB FedAvg algorithm. 

\begin{figure}[!ht]
  \centering
  \subfloat[Distribution of Model Update]{\includegraphics[width=0.24\textwidth]{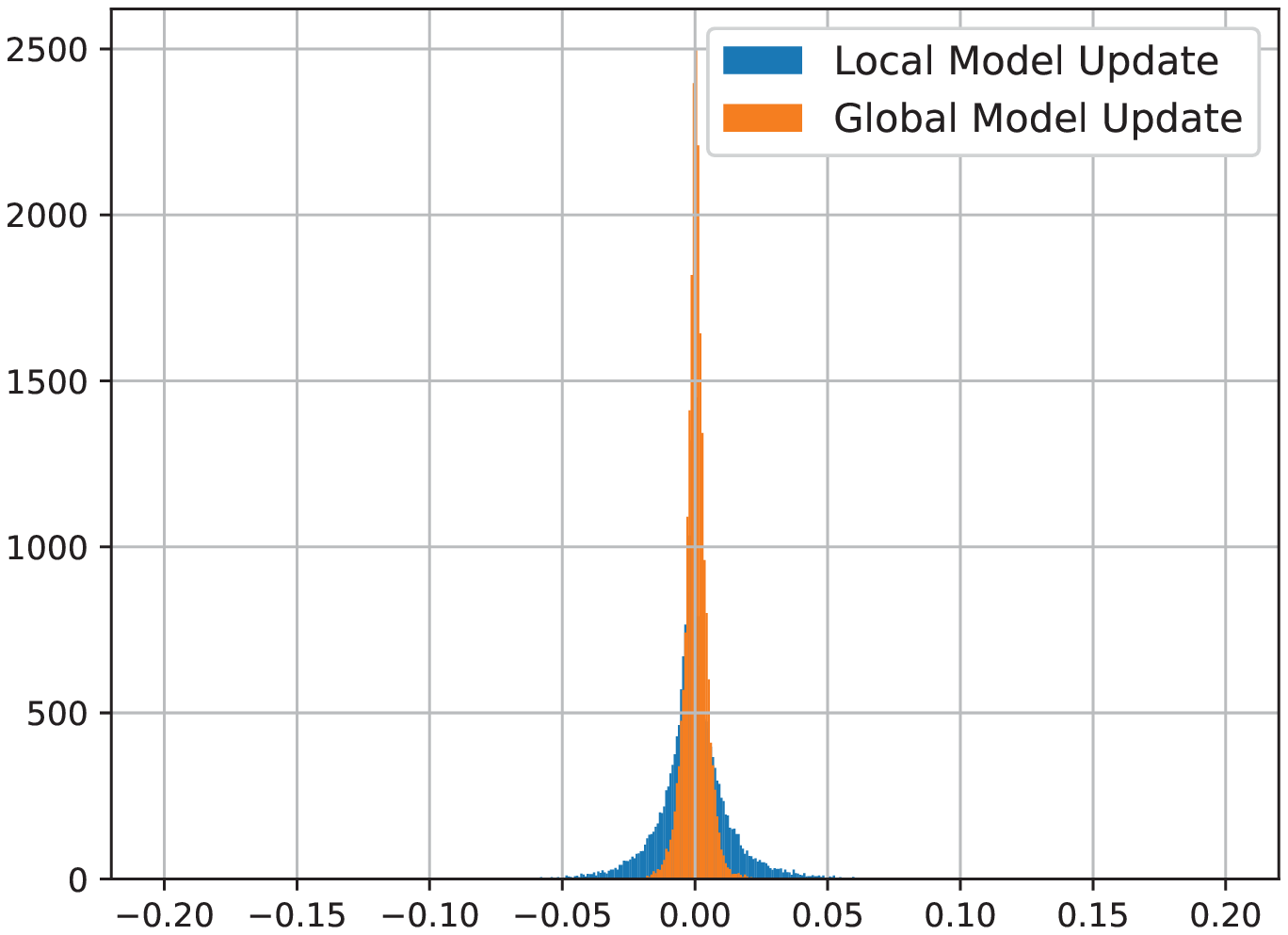}\label{fig:d1}}
  \hfill
  \subfloat[Distribution of Model]{\includegraphics[width=0.24\textwidth]{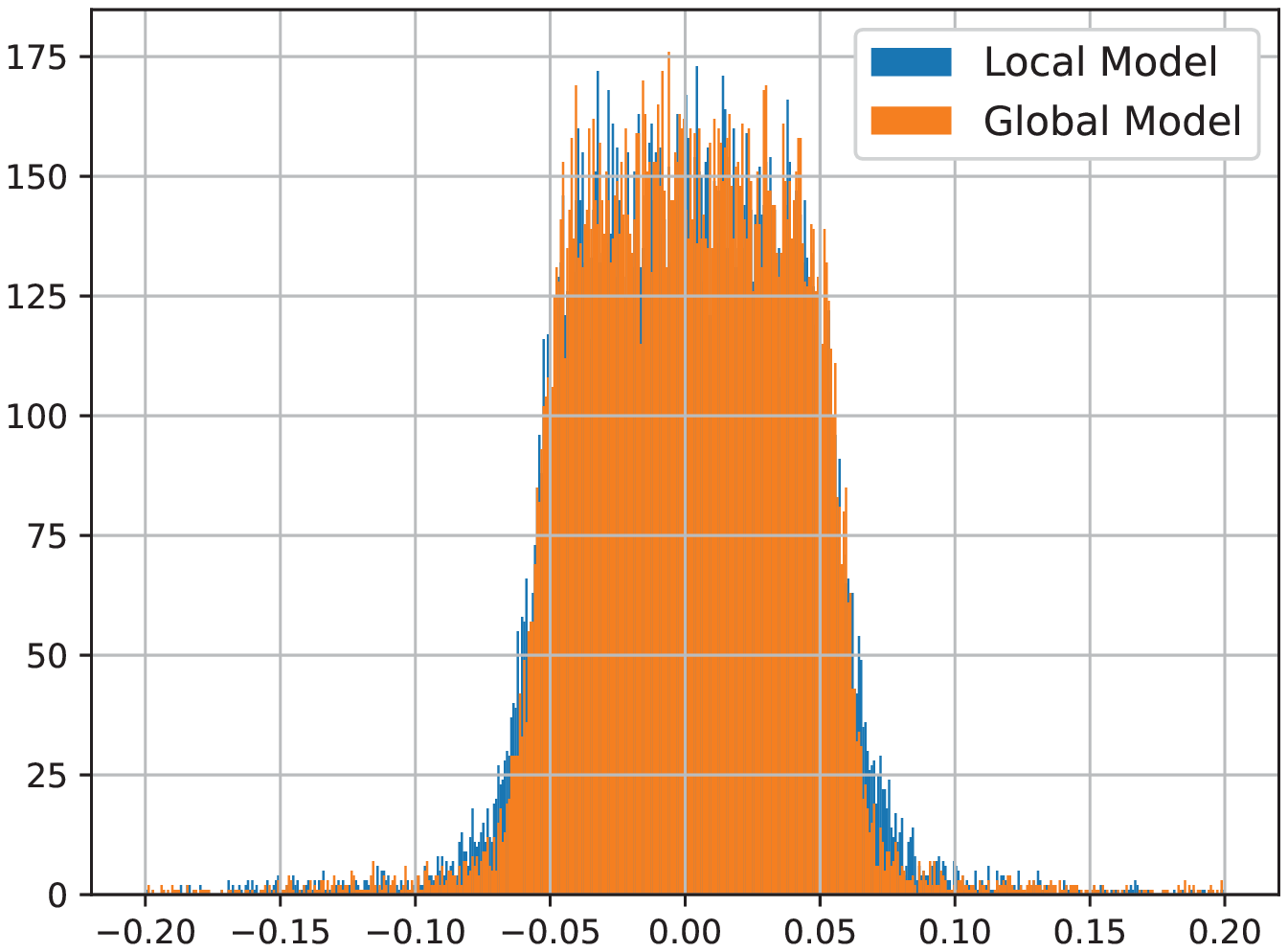}\label{fig:d2}}
  \caption{Distribution of Model Update and Distribution of Model with Non-IID data in FL}
  \label{distribution}
\end{figure}

To understand the difference between MB and MUB, we first derive the distributions of model and model update, respectively, in the non-IID data case. In Fig. \ref{distribution}, the distribution of model update in Fig. \ref{distribution}(\subref{fig:d1}) shows much less deviation than the model distribution in Fig. \ref{distribution}(\subref{fig:d2}). Further, in Fig. \ref{distribution}(\subref{fig:d1}), the distribution of local model update $u_t^k$ is similar to the distribution of global model update $u_t$. Similarly, in Fig. \ref{distribution}(\subref{fig:d2}), the distribution of local model $w_t^k$ is also close to the distribution of global model $w_t$. Thus the distribution of the local model and local model update is very close to its corresponding global distribution although the data is non-IID. 

In \cite{backdoor}, the author proposed the update norm clipping approach to ensure the norm of each model update is small enough so that the server is less susceptible to the poisoning models on backdoor attacks. This is proved to be a valid defense method for backdoor attacks without much impact on the performance of the main task. From the distributions of model and model update in Fig. \ref{distribution}, the $l_2-$norm of the model update distribution is much smaller than that of the model distribution. So the model update can naturally work like the norm-clipped model. And it should also apply to Byzantine attacks such as additive noise attacks and sign-flipping attacks.

\subsection{Initial Client Model Initialization}
In the classical MB FedAvg, the global model is first initialized by the server and then sent to LCs. Thus, each client has the same learning initialization point. Besides, the model of each client is accessible to the eavesdroppers during uploading. Thus the eavesdroppers have a chance to  perform membership inference attacks to infer the private data at LCs. To address that, ICMI is introduced as a new model initialization scheme. ICMI lets LC initialize its own model rather than using a common initial model sent by the server. SecAgg or AirComp can be further used to hide each LC's model from others, including the server and possible eavesdroppers, during the uploading stage.  The eavesdroppers can only obtain the aggregated model, which effectively prevents the membership inference attacks.

\subsection{Combined Scheme}
In the MUB scheme, since the model update rather than the model itself is shared by each client, the local model and the global model are effectively hidden from the eavesdroppers. However, due to the fact that the model initialization is done at the server and then is sent  to each client, the eavesdroppers may still be able to calculate the local model and the global model based on the initial global model and subsequent local model updates. To avoid that, MUB can be combined with ICMI to form the MUB-ICMI scheme, providing further privacy protection. With MUB-ICMI, no model information is made accessible to the eavesdroppers. Therefore, there is no need to use SecAgg or AirComp to hide the individual models. This simplifies the design of ICMI scheme by avoiding extra computation or communication overhead.

\section{Proposed scheme for FL Privacy \& security enhancement }
In the previous section, the motivation to apply the MUB scheme and ICMI scheme is articulated. In this section, the details of the proposed schemes are presented. 

\subsection{MUB FL}

\begin{figure}[!ht]
	\includegraphics[width=3.0in]{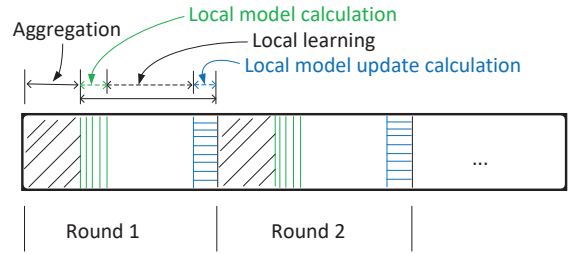}
	\centering
	\caption{Model Update Based FL Round}
	\label{fig:proposed_model}
	\centering
\end{figure}
As mentioned above, there are two stages for each FL round in classical MB-FL. The first stage is for model aggregation at the server, and the second stage is for the local learning at the LCs. For the MUB scheme, the aggregation stage aggregates model update rather than model, but the aggregation algorithm remains the same as the MB scheme. For the local learning part, there are three sub-stages taking place at each LC,  i.e., local model calculation, local learning, and local model update, as shown in Fig. (\ref{fig:proposed_model}). The process in each substage is summarized as follows.
\begin{itemize}
\item Local model calculation. When an LC receives the aggregated model update $u_t$, it first calculates the current local model $w_t^{k'}$ based on the previous local model $w_{t-1}^k$. The calculated local model, is $w_t^{k'} = w_{t-1}^k + u_t$. In the classical MB-FL, the local model before local learning is the same as the received global model $w_t$.
\item Local learning. The local learning is performed based on the local data to get the updated local model $w_t^k$, i.e., $w_t^k=w_t^{k'} - \eta \nabla F_k(w_t^{k'})$.
\item Local model update calculation. The local model update is calculated as $u_t^k = w_t^k - w_t^{k'}$.
\end{itemize}
The three substages can be summarized as:
\begin{subequations} \label{eq:three_stages}
\begin{eqnarray}
& & w_t^{k'} = w_{t-1}^k + u_t, \\ 
& & w_t^k = w_t^{k'} - \eta \nabla F_k(w_t^{k'}), \label{eq:local_com} \\ 
& & u_t^k = w_t^k - w_t^{k'}.
\end{eqnarray}
\end{subequations}
In ML, the gradient is usually defined as $g=\nabla F(w)$, so the local gradient in FL is $g_t^k = \nabla F_k(w_t)$, and the global gradient is $g_t=\sum_{k=1}^K\frac{|D_k|}{|D|}g_t^k$. Since $w_t^k=w_t^{k'}-\eta F_k(w_t)$ or $w_t^k=w_t^{k'}-\eta g_t^k$, we can rewrite $u_t^k= -\eta g_t^k$, i.e.,  the model update is related to the gradient. This can also be proved in equation (\ref{eq:local_model}), the model before local learning is $w_t$, so $u_t^k = -\eta \nabla F_k(w_t)=-\eta g_t^k$.  Specifically, this is for the scenario when only one local iteration is performed in one FL round on the local data. However, in FL, multiple local iterations might be performed to save communication bandwidth. When multiple iterations are executed, equation (\ref{eq:local_com}) can be rewritten as  
\begin{subequations}
\begin{eqnarray}
& & w_t^k(0) = w_t^{k'}, \label{eq:local_com_rewrite1} \\
& & w_t^k(j+1) = w_t^k(j) - \eta \nabla F_k(w_t^k(j)). \label{eq:local_com_rewrite2}
\end{eqnarray}
\end{subequations}
The calculated local model $w_t^{k'}$ is set as the initial local learning point as in equation (\ref{eq:local_com_rewrite1}). The local learning can iterate multiple times in one FL round. The learned model serves as the learning starting point in the next iteration, as shown in equation (\ref{eq:local_com_rewrite2}). When a total of $N$ ($N>1$) iterations are performed, the local model becomes $w_t^k(N)$. From this perspective, the MUB-FL is different from the existing  gradient based FL ($N=1) $ \cite{fedsgd}.

In the classical MB FedAvg, an LC only needs to perform local learning in each FL round. In MUB-FL, a client needs to execute the three substages, as shown in Fig. \ref{fig:proposed_model}. This might result in different local models and  different convergence behavior. When the model is initialized from the server and sent to LCs, the initial global model $w_1$ in the classical MB-FL is the same for all the clients. The local model after learning is $w_1^k = w_1 - \eta \nabla F_k(w_1)$. For MUB-FL,  with the initial local model $w_0^k=w_1$ and the initial global model update $u_1 = 0$,  the local model in the first round is $w_1^k = w_1^{k'} - \eta \nabla F_k(w_1^{k'})$, where $w_1^{k'} = w_0^k + u_1$. The local model after the first-round learning is the same as in the classical MB FedAvg. However, starting from the second FL round, the local model before learning $w_t$ for the classical MB-FL is the same for all LCs. The global model update $u_t$ for each LC in MUB-FL is still the same. However, since the local model $w_{t-1}^k$ from the previous round differs among different clients, the local model before learning $w_t^{k'}$ also differs across various clients. This is the main difference between classical MB-FL and MUB-FL. Although the local model before learning for clients is different in MUB-FL, the learning accuracy still converges to the same level as in classical MB-FL with no attacks. This is  verified in  the simulation results.

\subsection{ICMI FL}

ICMI aims to provide further privacy protection by hiding the individual model of each client. It initializes the model at each client rather than initializing the model at the server. So the initial models are different across different clients. Furthermore, SecAgg or AirComp can be used for model aggregation to further hide the individual client model. Several SecAgg protocols were proposed in \cite{secAgg}, e.g.,  masking with one-time pads, dropped user recovery using secret sharing, exchanging secrets efficiently, and minimizing thrust in practice. As the aggregated model is hidden from eavesdroppers and other third parties, the membership inference attacks can not be executed. Since only the initial model $w_1$ is different across LCs while the rest part keeps the same as in the classical  MB FedAvg, ICMI should also converge to a similar level as in classical MB-FL. This is verified  in the simulation.

\subsection{MUB-ICMI FL}
To enhance both security and privacy in FL, MUB and ICMI can be combined to form MUB-ICMI, where the model initialization is taken place at LCs, and only the model update (not the model itself) is uploaded. The MUB-ICMI algorithm is summarized in Algorithm \ref{alg:combined}.

\begin{algorithm}[]
\caption{MUB-ICMI FedAvg}
\begin{algorithmic}[1]
\STATE Each client initializes $w_0^k$, server initializes $u_1=0$
\STATE \textbf{Server executes:}
\begin{ALC@g}
\FOR{each round t=1,2,...}
\STATE $m \leftarrow \max(C\cdot K, 1)$
\STATE $S_t \leftarrow$ (random set of $m$ clients)
\FOR{each client $k \in S_t$ \textbf{in parallel}}
\STATE $u^k_t \leftarrow$ ClientUpdate($k$, $u_t$)
\ENDFOR
\STATE $u_{t+1} \leftarrow \sum_{k=1}^K \frac{|D_k|}{|D|}u_t^k$
\ENDFOR
\end{ALC@g}

\STATE \textbf{ClientUpdate($k$, $u_t$)}: // \textit{Run on client k}
\begin{ALC@g}
\STATE $\mathcal{B} \leftarrow$ (split $\mathcal{P}_k$ into batches of size $B$)
\STATE $w_t^{k'} \leftarrow w_{t-1}^k + u_t$
\STATE $w_t^k(0) \leftarrow w_t^{k'}$
\FOR{each local iteration $j$ from 0 to $N-1$}
\FOR{batch $b \in \mathcal{B}$}
\STATE $w_t^k(j+1) \leftarrow w_t^{k'}(j) - \eta \nabla F_k(w_t^{k'})(j)$
\ENDFOR
\ENDFOR
\STATE $u_t^k \leftarrow w_t^k(N) - w_t^{k'}$
\STATE return $u_t^k$ to server
\end{ALC@g}
\end{algorithmic}
\label{alg:combined}
\end{algorithm}

\section{Simulation results}
In this section, we first show the convergence of the proposed three schemes as well as the classical MB FedAvg algorithm by using image classification tasks under no attacks. Then the testing results of four different schemes under two different Byzantine attacks, i.e., additive noise attacks and sign-flipping attacks are presented. Compared with the classical MB-FL scheme, MUB-ICMI scheme is effective in defending against Byzantine attacks while still achieving good performance.

We consider a typical FL setting in the simulation where multiple clients are connected to the PS. Here, we use $K=100$ and $C=100\%$, that is, $100$ clients connected to the PS. And all of them participate in the ML tasks in each FL round. The image classification tasks are explored, and MNIST dataset is used. To present a convincing case, both multi-layer perception (MLP) and convolutional neural network (CNN) ML models are considered.  Different data distributions (both IID and non-IID) are exploited in the experiment. The MNIST dataset is a large dataset consisting of handwritten digits with digits $0-9$. It contains $60,000$ images for training and $10,000$ images for testing. Each image is formatted as $28\times28$ pixels. For the MLP model, only one hidden layer is used. For the CNN model, two convolutional layers are followed by the pooling layer with two fully-connected layers at the end. Since the common features of the images are in the same square or rectangular blocks, CNN usually achieves better performance than the MLP model. With IID data distribution, images are selected randomly and are allocated equally to each client. Non-IID data distribution allocates the images to the clients based on their labels. Each client is assigned two labels or digits. And each client is assigned around $600$ images for training. The testing is performed using the global model after aggregation in each round on the whole testing dataset. In the MUB-FL, the global model is calculated by accumulating the global model update starting from the initial global model. The learning model hyperparameters are learning rate size $\eta=0.01$, batch size $B=5$, local iteration count $N=2$.

\begin{figure}[!ht]
	\includegraphics[width=2.9in]{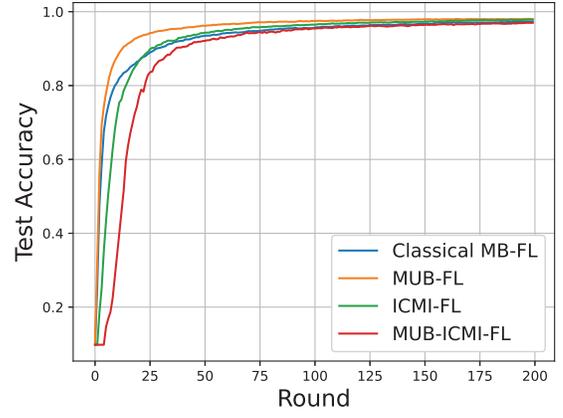}
	\centering
	\caption{Test accuracy of Non-IID data with CNN model without any attacks}
	\label{fig:convergence}
	\centering
\end{figure}

First, the convergence of the classical MB-FL and the proposed MUB-FL and ICMI-FL are demonstrated using MLP or CNN model under IID or non-IID data distributions without any attacks. Fig. \ref{fig:convergence} shows the testing accuracy of MNIST under a non-IID data distribution with the CNN model. All four algorithms converge after $200$ training rounds and also converge to the levels that are very close to each other. The MUB algorithm even achieves slightly better performance than the classical MB-FL during the learning process. The heterogeneity of data distribution has less impact when the model update rather than the model is aggregated. Under IID data distribution, MUB-FL achieves similar performance compared with the classical MB-FL. The MUB-ICMI algorithm performs slightly worse than the other three algorithms. 

To demonstrate the effectiveness of the proposed algorithms in defending against Byzantine attacks, we evaluate the proposed algorithms in several scenarios. First, we assume that  $20\%$, $30\%$, or $40\%$ of the clients are attackers. Two types of attacks are considered, additive noise attacks and sign-flipping attacks. In the additive noise attack, malicious clients add Gaussian noise to their local model updates and send them to PS. The malicious clients have the desire to add the noise with significant power. However, it is easy to detect it by computing the $l_2$-norm of the model update. In a sign-flipping attack, the malicious client flips the signs of the model updates while keeping the magnitude unchanged. So this attack is more brutal to be identified and thus more harmful to the system performance.
\begin{figure}[!ht]
  \centering
  \subfloat[IID data with Additive noise attacks]{\includegraphics[width=0.24\textwidth]{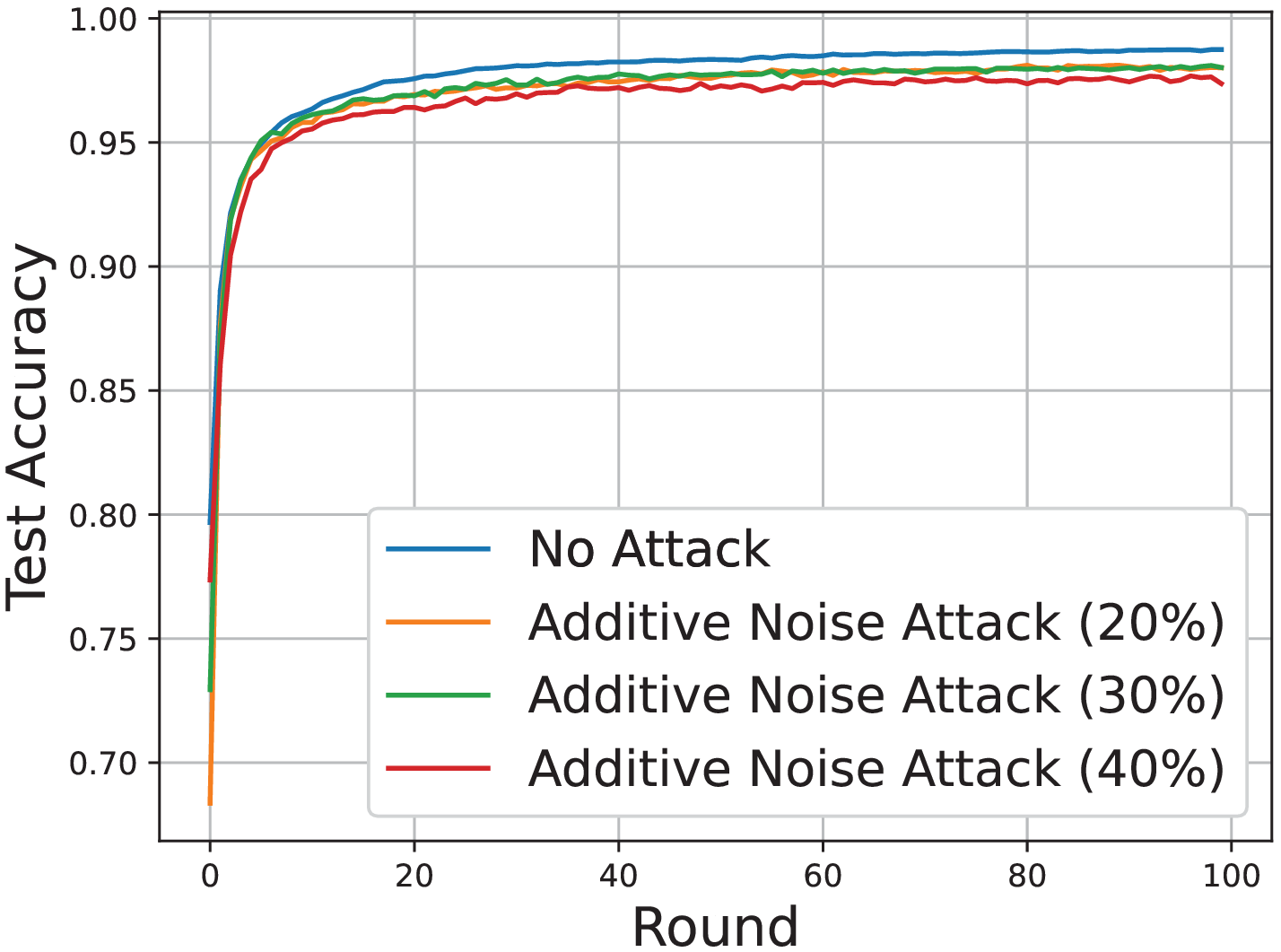}\label{fig:f1}}
  \hfill
  \subfloat[Non-IID data with Sign-flipping attacks]{\includegraphics[width=0.24\textwidth]{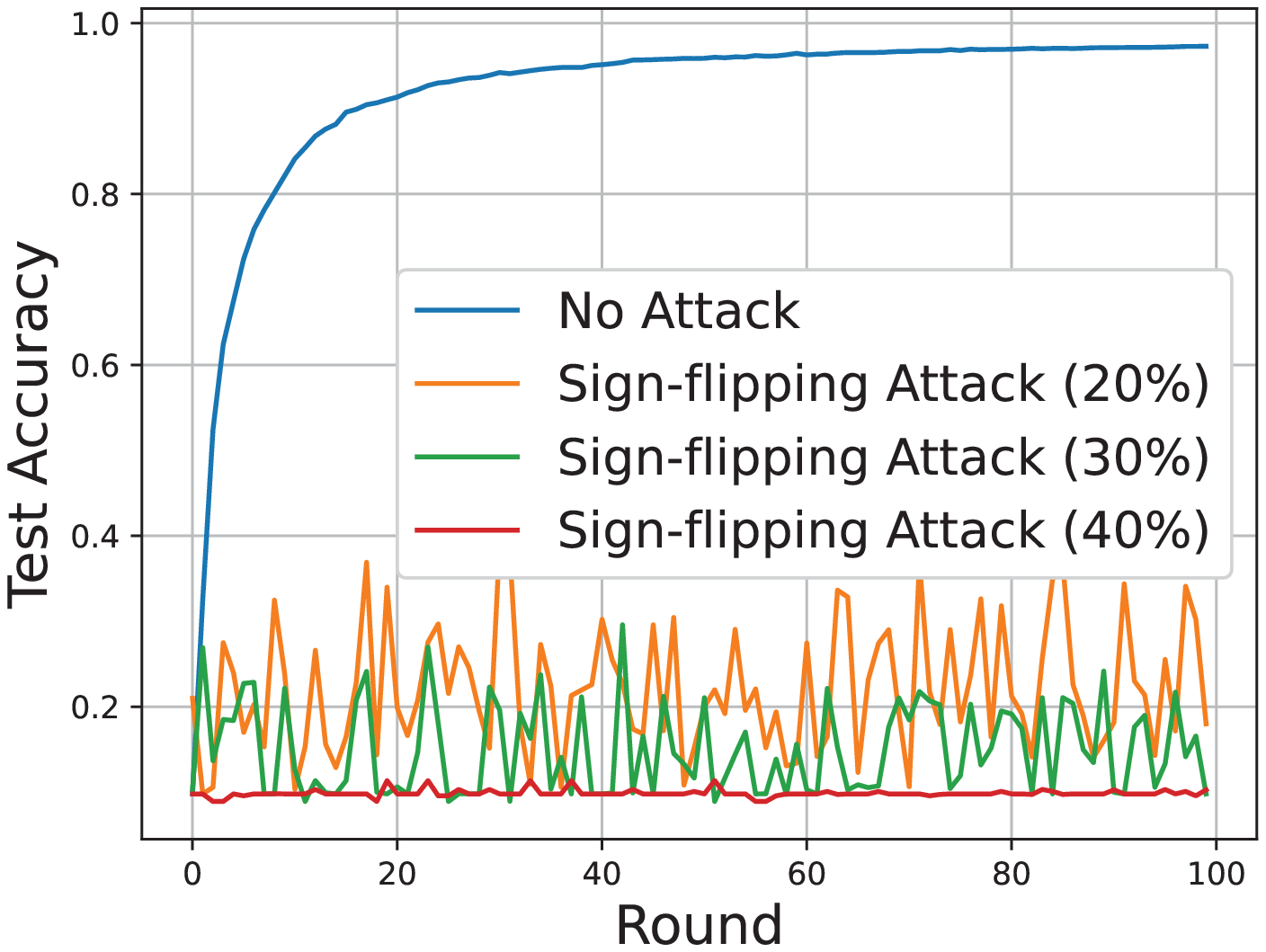}\label{fig:f2}}
  \caption{Test accuracy with CNN model using classical MB-FL}
  \label{classical_fedavg}
\end{figure}

Fig. \ref{classical_fedavg} presents the results of classical MB-FL with IID data under additive noise attacks as well as with non-IID data under sign-flipping attacks in the CNN model. When more malicious clients participate in model sharing, the performance becomes worse. The IID data distribution makes the test accuracy smoother under Byzantine attacks shown in Fig. \ref{classical_fedavg}(\subref{fig:f1}). Due to the non-IID data distribution, the test accuracy results in Fig. \ref{classical_fedavg}(\subref{fig:f2}) experience fluctuation during the learning. Compared with the additive noise attacks, sign-flipping attacks suffers much worse performance. When $40\%$ of the clients are malicious clients, the model almost learns nothing under the sign-flipping attacks. As shown in Fig. \ref{classical_fedavg}, the non-IID data distribution is more vulnerable to sign-flipping attacks. So in the following result, only the testing result of non-IID data distribution under sign-flipping attacks will be presented to demonstrate the effectiveness of the proposed schemes.

\begin{figure}[!ht]
	\includegraphics[width=2.8in]{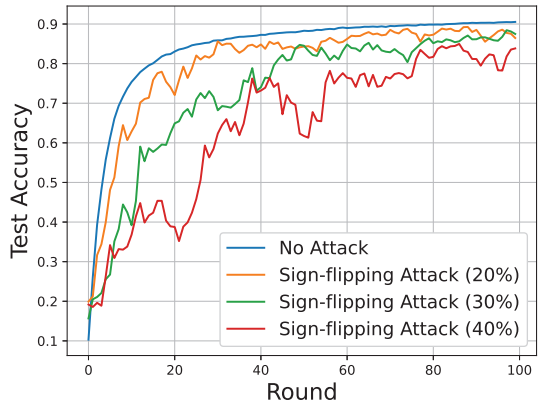}
	\centering
	\caption{Test accuracy of Non-IID data with MLP model using MUB scheme}
	\label{fig:model_update}
	\centering
\end{figure}
\begin{figure}[!ht]
	\includegraphics[width=2.8in]{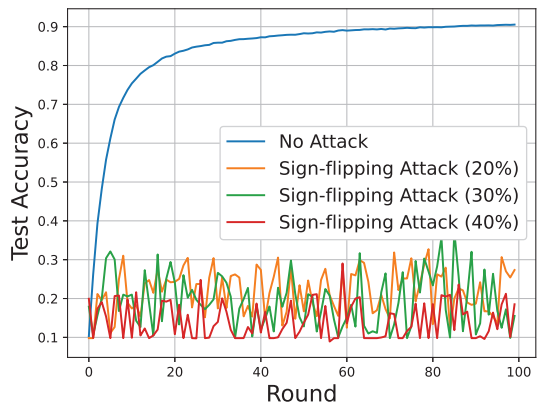}
	\centering
	\caption{Test accuracy of Non-IID data with MLP model using ICMI scheme}
	\label{fig:initialization}
	\centering
\end{figure}

Fig. \ref{fig:model_update} provides the results of non-IID data with MLP model using MUB scheme under sign-flipping attacks, and Fig. \ref{fig:initialization} presents the results of ICMI scheme under sign-flipping attacks. From Fig. \ref{fig:model_update}, we know MUB-FL can significantly defend against the sign-flipping attacks. After $100$ learning rounds, even in the worst case with $40\%$ malicious clients, the testing accuracy is still very close to the result without any attacks when it converges. Since the ICMI scheme is designed to enhance privacy, it does not help to defend the Byzantine attacks.  Thus the testing results in Fig. \ref{fig:initialization} are similar to the results with the classical MB-FL algorithm shown in Fig. \ref{classical_fedavg}(\subref{fig:f2}).

\begin{figure}[!ht]
	\includegraphics[width=2.8in]{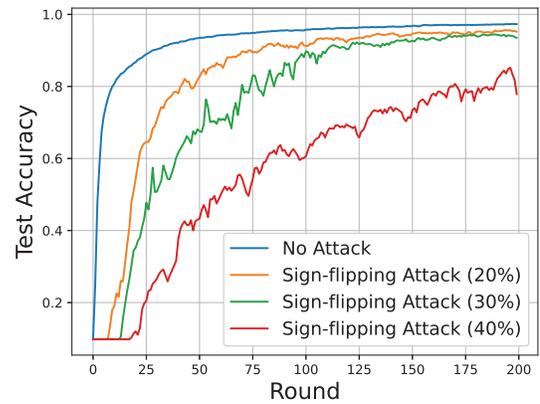}
	\centering
	\caption{Test accuracy of Non-IID data with CNN model using MUB-ICMI scheme}
	\label{fig:combined}
	\centering
\end{figure}

Finally, the MUB-ICMI scheme is applied to non-IID data distribution with the CNN model under sign-flipping attacks. To demonstrate the effectiveness of the algorithm, $200$ training rounds are executed. In Fig. \ref{fig:combined}, the scenarios with $20\%$ and $30\%$ malicious clients achieve similar results to the ``No Attack" case after $200$ training rounds. For the attack with $40\%$ malicious clients, although the performance is worse than the ``no attack" scenario as expected, it is still much better than the classical MB-FL shown in Fig. \ref{classical_fedavg}(\subref{fig:f2}).

\section{Conclusions}
In this paper, we proposed a new method using the MUB scheme in FL to defend against Byzantine attacks and the ICMI scheme to enhance privacy. The combined new MUB-ICMI can effectively improve both privacy and security in FL. Two types of Byzantine attacks were used to demonstrate the effectiveness of the proposed schemes. The convergence and effectiveness of the methods were presented using the MNIST dataset with both IID and non-IID data distributions. The theoretical analysis of the proposed algorithms and more simulation results on other datasets will be provided in the future.

\vspace{12pt}
\end{document}